\documentclass[12pt,preprint]{aastex}

\slugcomment{}

\shorttitle{CMB maps lensed by nonlinear structures}
\shortauthors{L. Ant\'on et al.}

\begin{document}


\title{CMB maps lensed by cosmological structures: \\
    Simulations and statistical analysis}


\author{L. Ant\'on, P. Cerd\'a-Dur\'an, V. Quilis and D. S\'aez}
\affil{Departamento de Astronom\'{\i}a y Astrof\'{\i}sica, 
       Universidad de Valencia, 46100, Burjassot, Valencia, Spain}
\email{diego.saez@uv.es}







\begin{abstract}
A method for ray-tracing through n-body simulations has been 
recently proposed.
It is based on a periodic universe covered by 
simulation boxes. Photons move along appropriate directions 
to avoid periodicity effects.
Here, an improved version of this method is applied to simulate  
lensed CMB maps and maps of lens deformations.
Particle mesh
n-body simulations with appropriate boxes and resolutions are used
to evolve the nonlinear inhomogeneities until present time.   
The resulting maps are statistically analyzed to look for
deviations from Gaussianity. These deviations are measured
--for the first time--
using correlations for configurations 
of $n$ directions 
($n \leq 6$). A wide range of angular scales are considered.
Some interesting prospects are outlined.

\end{abstract}



\keywords{cosmic microwave background --- cosmology:
theory --- large scale structure of universe},

\section{Introduction}

In this paper, lensed and unlensed maps of the 
Cosmic Microwave Background (CMB) temperature distribution 
are simulated and analyzed.
We are particularly interested in estimating
deviations with respect to Gaussianity. Any calculation in this field
requires the previous assumption of both a background 
universe and a distribution of cosmological inhomogeneities.
Taking into account results from the analysis of the WMAP 
(Wilkinson Microwave Anisotropy Probe) first year data \citep{ben03},
the cosmological background is assumed to be an inflationary 
cold dark matter universe with cosmological constant
and reionization. The reduced Hubble constant is
$h=10^{-2}H_{0}=0.71$ ($H_{0}$ being the 
Hubble constant in units of $Km \ s^{-1} Mpc^{-1}$), the
density parameters corresponding to baryons,  dark matter, and  
vacuum, are  $\Omega_{b} = 0.04$, $\Omega_{d} =0.23$ and  
$\Omega_{\Lambda} = 0.73$,        
respectively,
the total density parameter is
$\Omega = \Omega_{b} + \Omega_{d} + \Omega_{\Lambda} = 1$
(inflationary flat universe), and
the matter density parameter is 
$\Omega_{m} = \Omega_{b} + \Omega_{d} =0.27$.
A total reionization is assumed at redshift $z = 17$.
Cosmological tensor perturbations are absent, whereas the scalar
ones are adiabatic energy fluctuations with a Zel'dovich
spectrum and a Gaussian distribution. Under all these assumptions,
the CMBFAST code \citep{sel96} has been used to get the angular 
power spectrum of the CMB in the absence of lensing
(primary anisotropy plus the integrated Sachs-Wolfe effect
and reionization imprints; hereafter dominant anisotropy) 
and then, a method based on the Fast Fourier Transform 
\citep{sae96} is used
to simulate unlensed Gaussian maps of the temperature contrast
$\Delta_{_{D}}$ corresponding to dominant anisotropy. These
maps do not appear to be fully Gaussian because they are 
generated and analyzed by means of discretized numerical processes.
Deviations from Gaussianity are studied below in an appropriate 
interval of small angular scales. 
The simulated maps of dominant anisotropy are deformed by
lensing.

The CMB is lensed by cosmological 
objects evolving in the linear, mildly nonlinear, and strongly 
nonlinear regimes. All these structures deviate the propagation 
direction of the 
CMB photons. The resulting deviations deform the unlensed maps to
produce lensed ones. 
Linear and mildly
nonlinear structures can be studied by using analytical or 
semi-analytical methods and, consequently, lens deformations 
caused by this kind of inhomogeneities can be calculated without 
numerical n-body simulations of structure formation; however,
the lens effect of strongly nonlinear structures, as galaxy clusters
and substructures, is often studied by using ray-tracing 
through n-body simulations.
The history of the researches about lensing and its status at the end of
the last decade can be found in \citet{jai00}; in that paper, the
authors gave very exhaustive bibliography and
described and used the classic ray-tracing method based on 
multiple plane projections. Afterwards, another interesting method
was proposed by \citet{whi01}. This second method is based on many 
independent simulations, with 
different boxes and resolutions, tiling the photon trajectories. 
Finally, a third method using an unique simulation has been 
recently proposed \citep{cer04}. It is used here after
some improvements. The 
main features and advantages of this last method are now pointed out.

\begin{figure}
\epsscale{.80}
\plotone{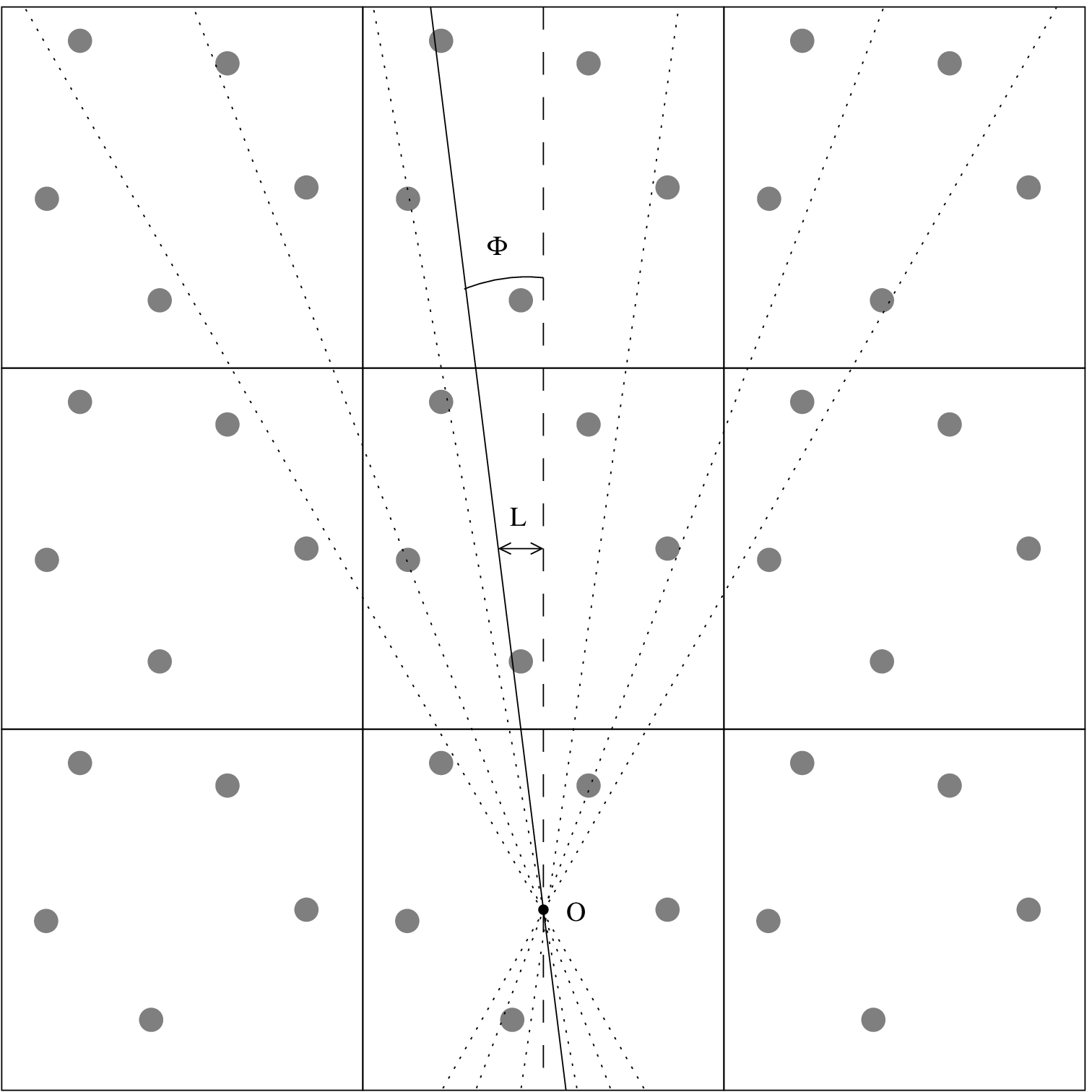}
\caption{Sketch of a 2D universe covered by squared patches.
Circles are structures, the observer is located at point 
$O$ and lines are photon trajectories 
\label{fig1}}
\end{figure}

The use of a n-body simulation constraint us to work in a periodic
universe tiled by identical boxes. Photons move in this unrealistic
periodic universe and the effect of periodicity is important. It 
magnifies lens deviations. This magnification depends on
the observation direction and, fortunately, there are directions leading
to negligible periodicity effects. 
They are hereafter called {\em preferred  
directions}. In order to understand the existence of these directions 
(minimizing periodicity effects) let us consider the two-dimensional (2D) 
sketch displayed in Fig.~\ref{fig1}, where
squares tile a bi-dimensional universe, O is the observer, 
circles are clusters, and lines are photon trajectories. If photons move 
parallel to the square edges (dashed line), they enter neighbouring squares 
through the same point and pass near
the same structures in all the squares; hence, lens deviations 
caused by repeated structures add and, consequently, 
a wrong magnified accumulative lens effect arises.
Now, the question is: what happens if photons follow the 
solid line? In such a case, 
photons enter successive boxes through different points 
separated a distance $L$
as that represented in Fig.~\ref{fig1}. For appropriate values of the 
angle $\phi$, distance
$L$ becomes greater than: (i) the distance at which clusters produce relevant
deviations of the photon directions (a few Megaparsecs), and (ii) the 
spatial scales of significant cluster correlations ($\sim 30 \ Mpc$); 
hence,  
distances $L > 30 \ Mpc$ suffice to ensure that
photons are deviated by distinct independent clusters located in
different uncorrelated regions. 
Since there are structures more extended than clusters 
in the n-body simulation,
a cutoff avoiding all the spatial scales larger than
a scale $L_{cut} \simeq L$ is performed 
in the peculiar potential given by the n-body 
simulation. It is worthwhile to notice that the simulation uses 
all the involved scales to calculate the gravitational potential
and to evolve particles.
The scales greater than $L_{cut} $ are subtracted from the
resulting gravitational potential only to estimate 
lensing. Using the proposed cutoff, CMB photons are lensed 
by distinct structures in successive boxes and, furthermore,
too  large scales which are not well described in our
simulation box are eliminated.
Finally, it is worthwhile to remark that, as a 
result of periodicity, there are no discontinuities in the points
where photons pass from a box to the next one (hereafter 
called {\em crossing points}); as discussed in the next paragraph,
this fact can be important
in order to get a good estimate of lens deviations. 

Roto-translations of the squares could be used to avoid the effects
of periodicity. If photons move through the dashed line (parallel to the 
square edges) but a translation of length $L$ is performed when photons arrive 
to the crossing points; then, photons are 
lensed by different clusters in distinct boxes (as it occurs for the
preferred directions in our procedure); nevertheless, roto-translations
produce unavoidable discontinuities in the crossing points. It is due 
to the fact that
the photon arrives to an edge at a certain point P, but it enters the next
square --after translation-- through a different point Q. 
The same occurs for rotations and 
roto-translations and, also, when we use the tiling method proposed by
\citet{whi01}, in which, different simulations are performed in each
square. These discontinuities are numerical artifacts and its importance 
cannot be neglected {\em a priori}.

If the 3D case is considered, the situation is 
analogous; then, our universe is covered by boxes and the trajectories 
parallel to edges are not appropriate to estimate true lens deviations
minimizing periodicity effects;
nevertheless, if an appropriate observation direction is chosen
and a cutoff is consistently performed, a 
large number of boxes can be crossed through different 
independent regions without 
reentering in the zone where
the CMB photon moved initially. 
Let us discuss this fact in detail.
A Cartesian coordinate system is considered in this discussion; 
its origin $O$ is placed 
at the center of a box, and the 
axis are parallel to the box edges whose length is $L_{box}$.
The observer is at $O$ and $\vec {n}$ is the unit vector in the 
direction of a line of
sight. This vector can be written either in terms of spherical
coordinates $\theta$ and $\phi$ or in the form 
$\vec {n} = 
(\cos \alpha, \cos \beta, \cos \gamma )$, where 
$ \alpha$, $\beta$, and $\gamma $ are the angles formed by 
$\vec {n}$ and the x, y, and z-axis, respectively.
This second notation is the most appropriate for us.
Evidently,
if a CMB photon go through the x-axis a distance OX, then, it advances 
the distances $OY= (\cos \beta / \cos \alpha) OX $ and
$OZ= (\cos \gamma / \cos \alpha) OX $        
along the y and z-axis, respectively. Hence,
if the photon crosses $n$ boxes in the direction of the x-axis,
it crosses $m= (\cos \beta / \cos \alpha) n$ 
($p= (\cos \gamma / \cos \alpha) n$) in the direction $y$ ($z$).
On account of these considerations, preferred directions 
are easily obtained; as an example, for quantities $n=5$, 
$m=1$, $p=1 + (1/6)$, and $L=256 \ Mpc$, if
the photon crosses a
box face orthogonal to the x-axis at point $M$ 
and it crosses the successive parallel face
at point $N$, the distance MN in the y-z plane 
is $L = L_{box} [(m/n)^{2} + ((p-1)/n)^{2}]^{1/2} = 51.9 \ Mpc $ and,
furthermore, the photon only return to the starting point 
after crossing $n/(p-1) = 30$ boxes in the x direction; namely, 
after going through a comoving distance of
$\sim 8000 \ Mpc$ in the preferred direction, which is the distance 
from $z \sim 5.2  $ 
to $ z=0  $ in the background universe under consideration.
At redshift $z = 5.2  $ the box size ($256 \ Mpc$) subtends
an angle of $ \sim 1.83 $ degrees. Since the observer is located at
the center of a box, for a $ 1.83^{\circ} \times 1.83^{\circ} $ map, 
the initial photon positions at $z = 5.2  $ --involved in our
ray-tracing procedure-- are all located inside the same initial box
and on a sphere having a comoving radius of $\sim 8000 \ Mpc$. 
For greater maps,
initial positions cover various boxes and, consequently, 
the resulting maps would have the imprint of the spatial 
periodicity. That is true even 
if each photon crosses independent 
regions through a direction close to the preferred one. 
A large enough number of 
small $ 1.83^{\circ} \times 1.83^{\circ} $ maps allow us
a good estimate of correlations for angles smaller than $\sim 0.25^{\circ}$
($\ell$ greater than $\sim 800$).

Units are chosen in such a way that 
$c=8\pi G =1$, where $c$ is the
speed of light and $G$ the gravitation constant.
Whatever quantity "$A$" may be, $A_{_{L}}$ and $A_{0}$ stand for
the $A$ values on 
the last scattering surface and at present time,
respectively. The scale factor is $a(t)$, where $t$ is the
cosmological time, and its present
value, $a_{0}$, is assumed to be unity, which is always possible 
in flat universes.

\section{Formalism}

Since lensing produces changes in the propagation direction of
the CMB photons, the temperature contrast $\Delta$ observed in the 
$\vec {n}$ direction is the dominant temperature contrast $\Delta_{_{D}}$ 
corresponding
to another direction $\vec {n}_{0}$ ( $\Delta (\vec {n}) = 
\Delta_{_{D}} (\vec {n}_{0})$ ). The difference $\vec {\delta} =
\vec {n}_{0} - \vec {n}$ is the deviation due to lensing; hence, 
\begin{equation}
\Delta (\vec {n}) =
\Delta_{_{D}} ( \vec {n} + \vec {\delta} )  
\label{basmap}
\end{equation}
The deviation field due to cosmological
structures is \citep{selj96}:
\begin{equation}
\vec{\delta} = -2 \int_{\lambda_{_{L}}}^{\lambda_{0}} 
W(\lambda) \vec {\nabla}_{\bot } \phi \ d \lambda \ ,
\label{devi}
\end{equation}
where $\vec {\nabla}_{\bot } \phi = - \vec{n} \wedge \vec{n}
\wedge \vec{\nabla} \phi$
is the transverse gradient of the peculiar gravitational potential, and
$W(\lambda) = (\lambda_{_{L}} - \lambda)/ \lambda_{_{L}} $.
The variable $\lambda$ is 
\begin{equation} 
\lambda (a) = H_{0}^{-1} \int_{a}^{1} \frac {db} {(\Omega_{m0}b+
\Omega_{\Lambda} b^{4})^{1/2}} \ .
\end{equation}                
The integral 
in the r.h.s. of Eq. (\ref{devi})
is to be evaluated along the background null geodesics.

The peculiar potential $\phi$ involved in Eq. (\ref{devi}) can be decomposed
in the form $\phi = \phi_{1} + \phi_{2}$, where $\phi_{1}$ ($\phi_{2}$) 
is the potential
created by the scales smaller (greater) than $L_{cut}$. For appropriate 
values of the separation 
scale $L_{cut}$, potentials $\phi_{1}$ and $\phi_{2}$ are independent and,
furthermore, $\phi_{1}$ can be obtained with a n-body simulation
in a box having a size larger than $L_{cut}$ 
(a few hundreds of Megaparsecs), whereas the 
potential $\phi_{2}$ can be simulated in
a very big box (a few thousands of Megaparsecs) 
by using the potential of the linear approximation.  
There are two kinds of boxes tiling the universe, which are
crossed by the same null geodesics. $L_{cut} = 60 \ Mpc$ is a good choice 
as it is discussed below.
The total deviation given by Eq. (\ref{devi})
is the addition of the deviations corresponding to the independent
potentials $\phi_{1}$ and $\phi_{2}$.

In the potential approximation \citep{mar90},
which is valid for linear inhomogeneities 
no larger than the horizon scale and also for nonlinear
structures, function $\phi$ satisfies the 
equation:
\begin{equation}
\Delta \phi = \frac {1}{2} a^{2} (\rho - \rho_{m0}) \ ,
\label{cden}
\end{equation}
where $\rho_{m0} = \Omega_{m0} \rho_{crit}$ 
is the background energy density for matter.
In the case of clusters, substructures and scales smaller than $L_{cut} $,  
the peculiar
potential $\phi_{1}$ is given by an appropriate n-body simulation
--based on Eq. (\ref{cden})--
at each time step. As it is well known,
the potential $\phi_{2}$ created by linear
scales larger than $L_{cut} $ has a constant spatial profile
and it
is proportional the ratio $D_{1}(a)/a$, where $D_{1}(a)$ is the growing
mode of the scalar density fluctuations in the model under consideration
(see \citet{pee80} and also \citet{ful00}). 
At present time, the spatial scales producing significant
lens correlations at angular scales $\alpha \leq  0.25^{\circ}$
($\ell \geq 800$) are not larger than the horizon scale ($3000h^{-1} \
Mpc$) and, consequently, either Eq. (\ref{cden}) or its Fourier counterpart
can be used to get $\phi_{20}$; namely, to get 
the constant spatial $\phi_{2}$ profile.
For any of the above  
potentials, the integral (\ref{devi}) is
performed on the background null geodesics
\begin{equation} 
x^{i}=x^{i}_{_{O}} + \lambda (a) n^{i} \ ,
\label{ng}
\end{equation}
passing by the point $O$ where our observer is located and
having directions $\vec {n}$. 

The chosen  $\vec {n}$ directions point towards 
the region of the sky we are mapping and cover the map in an 
appropriate way. We always use a regular coverage. 
These directions define the map nodes where temperature
contrasts are calculated and, consequently, they fix 
the map resolution and its size.
Once the lines of sight are chosen,
unlensed CMB maps of dominant anisotropy 
$\Delta_{_{D}} (\vec {n})$ (hereafter 
U maps) are built up by using the method described 
by \citet{sae96} (see \S 1); hence,
the directions $\vec {n}$ point toward the nodes of the 
U maps. The lens deviations of the potentials 
$\phi_{1}$ and $\phi_{2}$ are calculated for
the same directions (map nodes) by means of Eq. (\ref{devi}).
These partial deviations are added to get the
total deviation, which is used to obtain the direction  
$\vec {n} + \vec {\delta}$; finally, the exact Eq. (\ref{basmap}) is 
used --without any approximation-- to calculate a lensed 
$\Delta (\vec {n})$ map (hereafter L map). Since the direction 
$\vec {n} + \vec {\delta}$ does not point --in general-- towards
a node of the U map, appropriate interpolation
methods are necessary to get the quantity
$\Delta_{_{D}} ( \vec {n} + \vec {\delta} )$ involved in 
Eq. (\ref{basmap}). The U maps to be lensed has a resolution
close to one arcminute ($\ell \sim 10000$) and, consequently, interpolations 
are performed on scales much smaller than the angular scales 
affected by lensing ($800 \leq \ell \leq 3000$, see below).
The high resolution of the U maps allow us the use of linear interpolation
(which does not produce deviations from Gaussianity), and also 
nonlinear interpolation based on splines, which would only
alter Gaussianity for $\ell \sim 10000$.
Finally, the map of the differences $\Delta (\vec {n}) - 
\Delta_{_{D}} (\vec {n})$ is also calculated (hereafter D map); it is the
difference (node by node) between a L maps and its associated U map; hence,
relation L=U+D has an evident meaning.

Either the n-body potential or the linear one can be used to calculate the integral
(\ref{devi}) in position space, but we can also work in momentum
space; in fact, as it was detailed in \citet{cer04}, after simple Fourier 
algebra, one easily get:
\begin{equation}
\vec{\delta }(\vec{x}_{_{O}},\vec{n}) =\frac {2i}{(2 \pi)^{3/2}}
\int  \frac {\vec{k}_{\bot }} {k^{2}}
F_{\vec {k}}(\vec {n}) e^{-i \vec {k} \vec {x}_{_{O}}} d^{3} k  \ ,
\label{funda1}
\end{equation}
where
\begin{equation}
F_{\vec {k}}(\vec {n}) = \int_{\lambda_{_{L}}}^{\lambda_{0}}
W(\lambda ) B(\lambda ) e^{-i \lambda \vec {k} \vec {n}}
\delta_{\vec {k}}(\lambda ) d \lambda  \ ,
\label{funda2}
\end{equation}
with $\vec{k}_{\bot} = \vec {k} - (\vec {n} \cdot \vec {k})
\vec {n}$ and $B = - \rho_{m0} / a$.

The n-body simulations are performed using a PM code \citep{hoc88} 
which was used, tested, and described in detail in \cite{qui98}. 
This code gives function $\delta_{\vec{k}}$ at each
time step and, then, for a given vector $\vec {n}$, 
the integral of the r.h.s. of 
Eq. (\ref{funda2}) can be performed to get function $F_{\vec {k}}(\vec {n})$.
Once this function has been calculated, Eq. (\ref{funda1}) is a 
Fourier transform which gives the deviation 
$\vec{\delta }(\vec{x}_{_{O}},\vec{n})$ corresponding to the
direction $\vec {n}$ and to the observer located at point 
$\vec{x}_{_{O}}$; hence, the deviations associated to all the observers
located in the nodes of the Fourier box in position space are
simultaneously calculated for a given direction $\vec {n}$.
This information is useful to estimate correlations as 
it was analyzed in \citet{cer04}. Similar procedures were
used by \citet{ali02} to study the Rees-Sciama effect in
momentum space. In the case of linear scales larger than 
$L_{cut}$, the values of $\delta_{\vec{k}}$ necessary to calculate 
the integral in 
Eq. (\ref{funda2}) are given by the linear
approximation to the evolution of density perturbations. 

\begin{figure}
\epsscale{.80}
\plotone{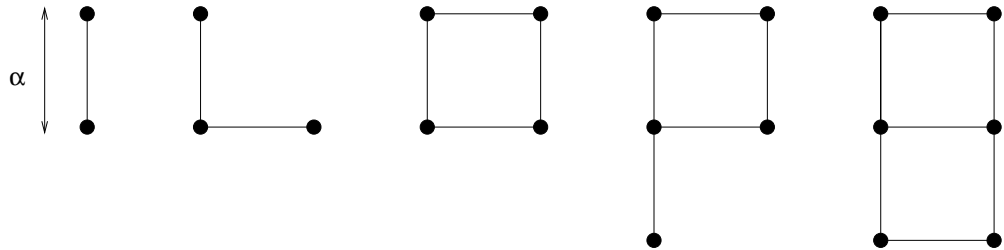}
\caption{Configurations of $n$ directions for $2 \leq n \leq 6$.
The sets of $n$ directions draw these Figures on the
Last Scattering Surface. 
\label{fig2}}
\end{figure}

Our U, L and D maps must be statistically analyzed. In 
order to do that, correlations associated to n directions (or
n points on the last  scattering surface) are estimated for
$n \leq 6$. Given a map of the variable $\zeta $ and $m$ directions, 
the angular correlations are:
\begin{equation}
C_{m} = \langle \zeta (\vec {n}_{1}) \zeta (\vec {n}_{2}) 
\cdots \zeta (\vec {n}_{m}) \rangle \ ,
\label{corr}
\end{equation}
where the average is over many maps.
Some of these averages are estimated,
in next section, for U, L and D maps and 
for $m \leq 6$.
Correlations depend on the relative positions of the 
chosen directions, namely, they depend on the figures that 
these directions draw on the last scattering surface. In this
paper we have used the sets of directions 
displayed in Fig.~\ref{fig2}, where 
the basic correlation angle, $\alpha$, is defined as
the angle formed by the two directions of the case $n=2$.
In Gaussian statistics it is well known that
\citep{pee80}: (i) all the 
$C_{m}$ correlations vanish for odd $m$ and, (ii)
for even $m$, all the correlations can be written in terms of 
$C_{2}$. Hereafter, these even Gaussian correlations 
are denoted $C_{gm}$ (the suffix g stands for Gaussian).
For the configurations of Fig.~\ref{fig2}, one easily 
obtains the relation:
\begin{equation}
C_{g4}(\alpha) = 2C^{2}_{2}(\alpha) + C^{2}_{2}(\sqrt{2} \alpha) \ ,
\label{cg4}
\end{equation}
leading to $C_{g4}(0) = 3C^{2}_{2}(0)$ for $\alpha = 0$, and the equation:                                              
\begin{eqnarray}
\nonumber
C_{g6}(\alpha) & = & 3C^{3}_{2}(\alpha) + 
2C_{2}(\alpha) C^{2}_{2}(\sqrt{2} \alpha)+
2C_{2}(\sqrt{5} \alpha) C^{2}_{2}(\alpha)+ \\
\nonumber 
& + & 4C_{2}(\alpha) C_{2}(2 \alpha) C_{2}(\sqrt{2} \alpha) 
+ 2C_{2}(\sqrt{5} \alpha) C^{2}_{2}(\sqrt{2} \alpha)+\\
& + & C_{2}(\alpha) C^{2}_{2}(\sqrt{5} \alpha)+
2C_{2}(\alpha) C^{2}_{2}(2 \alpha) \ , 
\label{cg6}
\end{eqnarray}
which takes on the form 
$C_{g6}(0) = 15C^{3}_{2}(0)$ for $\alpha = 0$.
Furthermore, according to \citet{ber97} and \citet{kes02}, if the
unlensed maps are assumed to be Gaussian, the non-Gaussian lensed ones
have vanishing $C_{m}$ correlations for odd $m$ values.
Our simulated U maps are almost Gaussian by construction, with small
deviations from Gaussianity associated to numerical generation;
hence, only these numerical deviations could lead to small  
odd correlations without any physical meaning. We are not interested in them;
Anyway, the correlations $C_{3}$ 
and $C_{5}$ have been estimated for the configurations of Fig. 2 and
the resulting values have appeared to be compatible with zero.
On account of these considerations, 
the simplest {\em method to estimate deviations from Gaussianity} 
is as follows: (a) extract 
the $C_{2}$, $C_{4}$, and $C_{6}$ correlations from the 
simulated maps, (b) calculate $C_{g4}$, and $C_{g6}$ functions 
from $C_{2}$ assuming Gaussianity, namely, using Eqs. 
(\ref{cg4})--(\ref{cg6}) and, (c) compare 
$C_{4}$ and $C_{6}$ correlations with functions 
$C_{g4}$ and $C_{g6}$, respectively. 

\begin{figure}
\epsscale{0.4}
\plotone{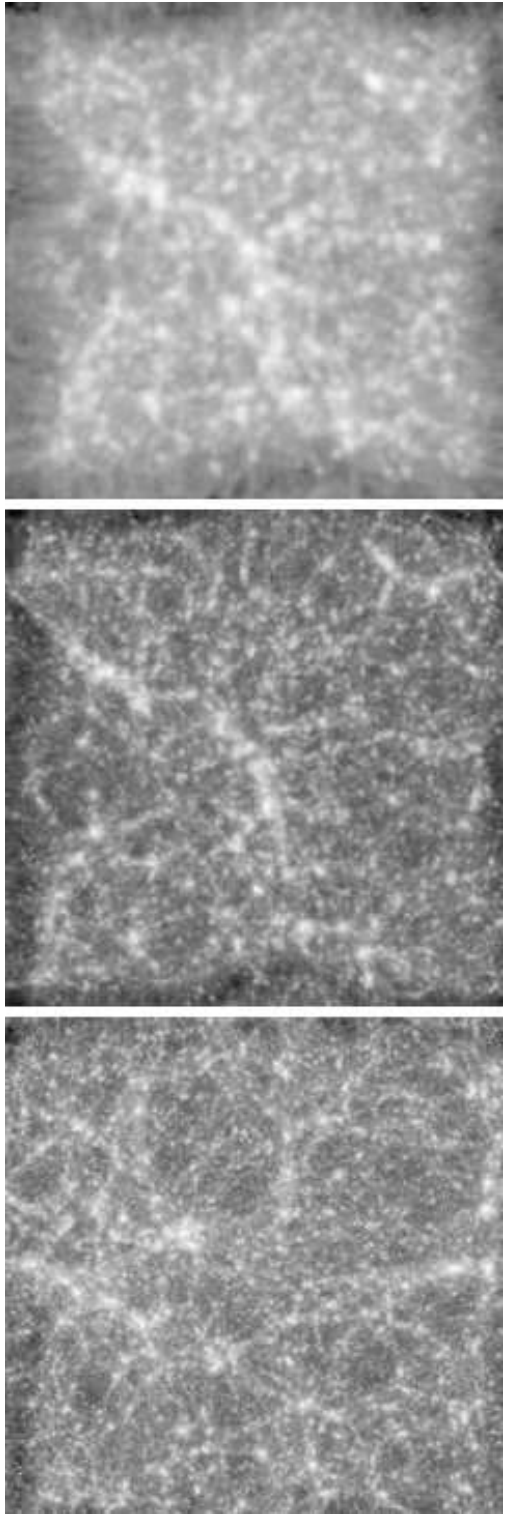}
\caption{integrated column density along an edge
of the simulation box. Top, middle and 
bottom panels correspond to a LR, IR and HR 
simulation, respectively. 
\label{fig3}}
\end{figure}

Now, let us describe the working method used to 
calculate lens deviations caused by the potential
$\phi_{1}$ (scales smaller than $L_{cut}$). 
First a preferred direction is fixed and a set of 
neighbouring directions are chosen to have a uniform pixelisation of
a small map of about $2^{\circ} \times 2^{\circ}$. This is the 
crucial point of the method. Afterwards, either Eq. (\ref{devi})
or Eqs. (\ref{funda1})--(\ref{funda2}) are solved to get deviations and,
finally, these deviations are used to deform U maps of dominant
anisotropy.
Now, let us look for appropriate values of the
box size and resolutions to be used in our PM n-body simulations
($\phi_{1}$ calculation).
The angular power spectrum of the lens deformations is
expected to be particularly relevant for $\ell > 800$; 
this means that correlations must be 
calculated for $\alpha < 0.25^{\circ}$ and, consequently,
the angular size of the simulated maps should be greater than 
$\sim 1^{\circ}$. Our map is chosen to be the image of a box face
orthogonal to the line of sight of its 
center and located at the initial redshift; with this choice, 
photon directions do not spread on many boxes at a given time, and 
there is no any undesirable effect produced by the 
spatial periodicity
of the fictitious universe under consideration.
For this type of maps,
the angular size depends on both $L_{box}$ and
$z_{in}$; it can be easily proved that, in our background universe, 
for $z_{in} = 5.2$ and 
$L_{box}=256 \ Mpc$, the resulting map has a large enough 
edge of 
$1.83^{\circ} $. Furthermore, as stated in \S 1,  
the separation distance $L$ between the regions crossed by                      
CMB photons in successive boxes is $51.9 \ Mpc $, the number of 
boxes crossed between redshifts 5.2 and 0. is $\sim 30$, and there
are hundreds of cluster in the box.          
Three types of simulations with different resolutions are
used in this paper: (i) Low Resolution (LR) simulations 
involving 128 cells per edge and $128^{3}$ particles,
(ii) Intermediate Resolution (IR) simulations 
using 256 cells per edge and $256^{3}$ particles, and
(iii) High Resolution (HR) simulations                   
with 512 cells per edge and $512^{3}$ particles. It is evident
that even our HR simulations,  with a cell size of $0.5 \ Mpc$
and an effective spatial resolution of two cells are not good
enough to describe the high density contrasts of core clusters;
nevertheless, as it will be discussed in detail in next section,
higher resolutions seem to be unnecessary to study the
lens deformations of the CMB maps.
Fig.~\ref{fig3} shows the column density for particular LR (top), 
IR (middle), and HR (bottom) simulations. Evidently, structures have 
developed in the three cases, but they have reached different levels 
of evolution. The diffuse figure of the top panel (LR simulation) 
indicates a low level of evolution (extended structures). The 
other panels exhibits a better definition because structures have 
evolved more.

A distance $L=51.9 \ Mpc$ suffices to deal with clusters and substructures
(see \S 1),
but a cutoff eliminating all the scales greater than 
a maximum one $L_{cut}$ (close to $L$ or smaller 
than it) is necessary to
ensure that photons cross different large scale structures in distinct 
boxes. Cutoffs at scales $L_{cut}$ of $60 \ Mpc$, $30 \ Mpc$, $15 \ Mpc$
and $7.5 \ Mpc$ are independently used in next section. 
A few words about these cutoffs
and their consequences are worthwhile.
The n-body simulations use all the scales corresponding to the chosen 
values of $L_{box} $ and $\Delta $ (spatial resolution); nevertheless,
once the n-body simulation has been developed up to a certain time, 
any of the mentioned cutoffs can be performed in the peculiar 
gravitational field responsible for the lens effect (all
the scales larger than $L_{cut}$ are thus erased). The n-body particles
evolve under the forces associated to the full $\phi $ field inside 
the box, whereas the
final potential --after the cutoff-- is only used to calculate lens
deviations using Eq. (\ref{devi})
(or Eqs. (\ref{funda1}) and (\ref{funda2})).

In order to perform a certain cutoff,
the density contrast, $\delta_{\vec {x}}$,                                 
and its Fourier transform, $\delta_{\vec {k}}$                                  
are calculated using  n-body simulations and, then                                                   
the Fourier transform of the total peculiar potential inside the box,                                 
$\phi_{\vec {k}}$, is calculated using the Fourier counterpart of           
Eq. (\ref{cden}): $\phi_{\vec {k}} \propto \delta_{\vec {k}}/k^{2}$. 
Afterwards, quantities $\phi_{\vec {k}} $ are avoided 
for any $k < 2 \pi / L_{cut}$ and, finally, an inverse
Fourier transform is used to get the peculiar potential $\phi_{1} $
after cutoff.
Scales of $60 \ Mpc$, $30 \ Mpc$ are larger than the scale 
$8h^{-1} \ Mpc \sim 11.3 \ Mpc$ and, consequently they
are linear, but the scales  $15 \ Mpc$ and $7.5 \ Mpc$ are
mildly nonlinear. All these scales are well described in a 
box having $256 \ Mpc$ per edge.

Any cutoff should produce two main effects: 
(i) it should erase too large 
spatial scales 
which are not well described in the n-body simulation, and (ii) 
the unaltered scales should be small enough to ensure that the CMB 
photons cross distinct structures in different boxes (along the
preferred directions). The erased scales are either linear or 
mildly nonlinear and, consequently,
they can be studied either using the linear approximation to
gravitational instability \citep{bar80} or applying 
appropriate techniques as the Zel'dovich approximation \citep{zel70}, 
the adhesion model (see \citet{san89} for a review), 
the lagrangian perturbation approach \citep{mou91}, and so on;
hence, lens deformations produced by scales greater than $L_{cut}$ 
can be calculated without n-body
simulations. Let us now consider the cutoff at $L_{cut} = 60 \ Mpc$.
Some scales smaller than $60 \ Mpc$ are nonlinear and couple
among them and with other scales, the n-body simulation is a good
description of the coupled evolution of all these scales. Once the lens 
effect produced by scales smaller than $60 \ Mpc$ has been estimated, 
the effect produced by strictly linear scales greater than $60 \ Mpc$
(complementary scales)
can be analytically or numerically calculated 
using the linear approximation and then, taking into account that linear
scales do not couple among them, and also that these large scales
are not expected to be significantly affected by the
nonlinear ones (which are much smaller), the resulting effect 
of the complementary scales can be added 
to the effect of scales smaller than $60 \ Mpc$ to get the total 
lens deformations. For 
$L_{cut}$ values much smaller than $60 \ Mpc$, e.g. $7.5 \ Mpc$,
all the complementary scales (greater than $7.5 \ Mpc$) are not
linear and, furthermore, nonlinear scales smaller and greater 
than $7.5 \ Mpc$ are coupled; these facts lead to a complicate
situation. By these reasons, highly linear $L_{cut}$ values, e.g. $60 \ Mpc$,
are preferable in order to get the total lens effect, but 
mildly nonlinear values as $7.5 \ Mpc$ can be used to get 
interesting information 
(see next section).  

In this paper, the lens effect produced by scales smaller 
than $L_{cut}$ is calculated from an 
initial redshift $z_{in} \sim 5.2$ to present time. This 
value of $z_{in} $ seems rather arbitrary, nevertheless, 
in next section it is proved that, for 
the scales under consideration, the lensing produced 
from decoupling to $z = 5.2$ can be neglected.

For $L_{box} = 128 \ Mpc$,
the size of the resulting maps is a little small ($\sim 0.9^{\circ}$), 
the number of crossed 
boxes seems to be too large ( $\sim 60 $), 
the $L $ value ( $\sim 25 \ Mpc$ ) is only marginally
admissible to deal with independent clusters in neighbouring boxes (see \S 1), 
and  
the cutoffs must be smaller than $\sim 25 \ Mpc$; hence, 
this box size (and similar ones)
seems to be a little small. There is only one advantage 
for $L_{box} = 128 \ Mpc$: that 
the cell size of the n-body simulations can be reduced
up to $0.25 \ Mpc$, with the same computational cost as in  
the HR simulations --in boxes of $256 \ Mpc$-- used in this paper.
In the case of large values $L_{box} \geq 256 \ Mpc$, a cell 
size as small as $0.25 \ Mpc$ requires too much memory; fortunately,
such a high resolution seems not to be necessary to deal with
CMB lensing (see \S 3). 
On account of these considerations, it seems that 
$L_{box} = 256 \ Mpc$ is a good choice to obtain appropriate maps allowing
statistical analysis.

Finally, a few words about the calculation of the deviations produced
by $\phi_{2}$; namely, by the scales greater than $L_{cut} = 60 \ Mpc$.
In this case, a big box of $5000 \ Mpc$ is considered with a resolution
of $\simÂ10 \ Mpc$ (512 nodes per edge). The potential $\phi_{2}$
created by the
linear scales greater than $60 \ Mpc$ is calculated, at present time, 
by using the Fourier
counterpart of Eq. (\ref{cden}) and the power spectrum $P(k) $ of the 
model under consideration (with the normalization $\sigma_{8} = 0.9$) and, 
then, this potential
is evolved taking into account that it is proportional to $D_{1}(a)/a$.
The lens effect produced by scales larger than $L_{cut} = 60 \ Mpc$ is
calculated from decoupling ($z_{in} \sim 1100$) to present time.
Three of these boxes cover all the photon trajectories from 
this initial redshift to present time. For an appropriate preferred direction,
the distance L is close to $2000 \ Mpc$ and,
consequently, we can say that, for the scales which produce 
significant contributions to the lens correlations we
are estimating ($\ell > 800$), 
photons cross different boxes through distinct 
linear inhomogeneities. In this case, there is no problem to get a map  
greater than those of $ 1.83^{\circ} \times 1.83^{\circ} $ associated to
the n-body simulations. We can easily obtain a map of 
deviations having a surface 
sixteen times greater (hereafter a $ 7.32^{\circ} \times 7.32^{\circ} $ map),
which can be superimposed to a mosaic of sixteen 
$ 1.83^{\circ} \times 1.83^{\circ} $ maps of $\phi_{1}$ deviations to get 
a large map of total deviations.

\section{Results}

Two types of maps (hereafter 1 and 2) have been created and analyzed. 
Type 1 maps are small simulations
of the partial lens effect produced by scales smaller than $L_{cut}$; among
them we can distinguish:
U (unlensed), L (lensed) and D (deformation) maps. All these maps have 
$128 \times 128$ nodes and $ 1.83^{\circ} \times 1.83^{\circ} $
(excepting two cases described below).
Lens deviations can be calculated with LR, IR and HR simulations and 
the associated maps are denoted LR-L, LR-D and so on; hence,
we have seven sets of maps: the first one is a set
of eight hundreds U maps to be deformed by lensing, 
each of the remaining sets (LR-L, LR-D, IR-L, IR-D, HR-L and HR-D)
contains sixteen thousands maps generated by deforming the 
U maps with the deviations of twenty
n-body simulations. Each of these seven sets has been analyzed to
get both the averaged angular power spectrum and 
the mean of the $C_{m}$ correlations for $2 \leq m \leq 6$ (see \S 2).
Type 2 maps correspond to the 
lens effect produced by all the scales.
Only U (unlensed) and L (lensed) maps (no D maps)
are considered in this case. All these maps have 
$512 \times 512$ nodes and $ 7.32^{\circ} \times 7.32^{\circ} $.
The U maps are lensed by 
total lens deviations, in other words, they are lensed by the
superimposition of a map of the deviations produced by structures 
with spatial
scales larger than $L_{cut}$, and a mosaic tiled by sixteen 
$ 1.83^{\circ} \times 1.83^{\circ} $ small maps of lens deviations 
produced by scales smaller than $L_{cut}$. The discontinuities of the
mosaic in the edges of the small maps do not significantly affect
the estimate of correlations for angular scales $\alpha < 0.25^{\circ}$,
which are much smaller than the size of these small maps.

L maps are deformations of the U maps and, consequently, 
neither L and U nor D and U  maps are statistically independent. 
According to the relation L=U+D (see \S 2), the following equation is 
satisfied:
\begin{equation}
\langle L,L \rangle - \langle U,U \rangle =
\langle D,D \rangle + 2 \langle D,U \rangle  \ .       
\label{cordep}
\end{equation}
The l.h.s of this equation is the difference between the 
$C_{2} $ correlations of the L and U maps. The corresponding 
$C_{\ell}$
quantities have a strongly oscillatory character as it
was emphasized  
in \citet{selj96} and \citet{hu00} (see also Fig.~\ref{fig5}).
The same methods used to get the oscillatory form of
the solid line of Fig.~\ref{fig5} (from L and U maps), have 
been also used to get the $C_{\ell}$ coefficients
of the D maps (associated to the first term of the r.h.s.).
The resulting coefficients
do not oscillate (see e.g. Fig.~\ref{fig9}).
The oscillations associated to the l.h.s. of Eq. (\ref{cordep})    
are due to the cross correlations
of D and U maps appearing in the second term of the r.h.s.

\begin{figure}
\epsscale{0.8}
\plotone{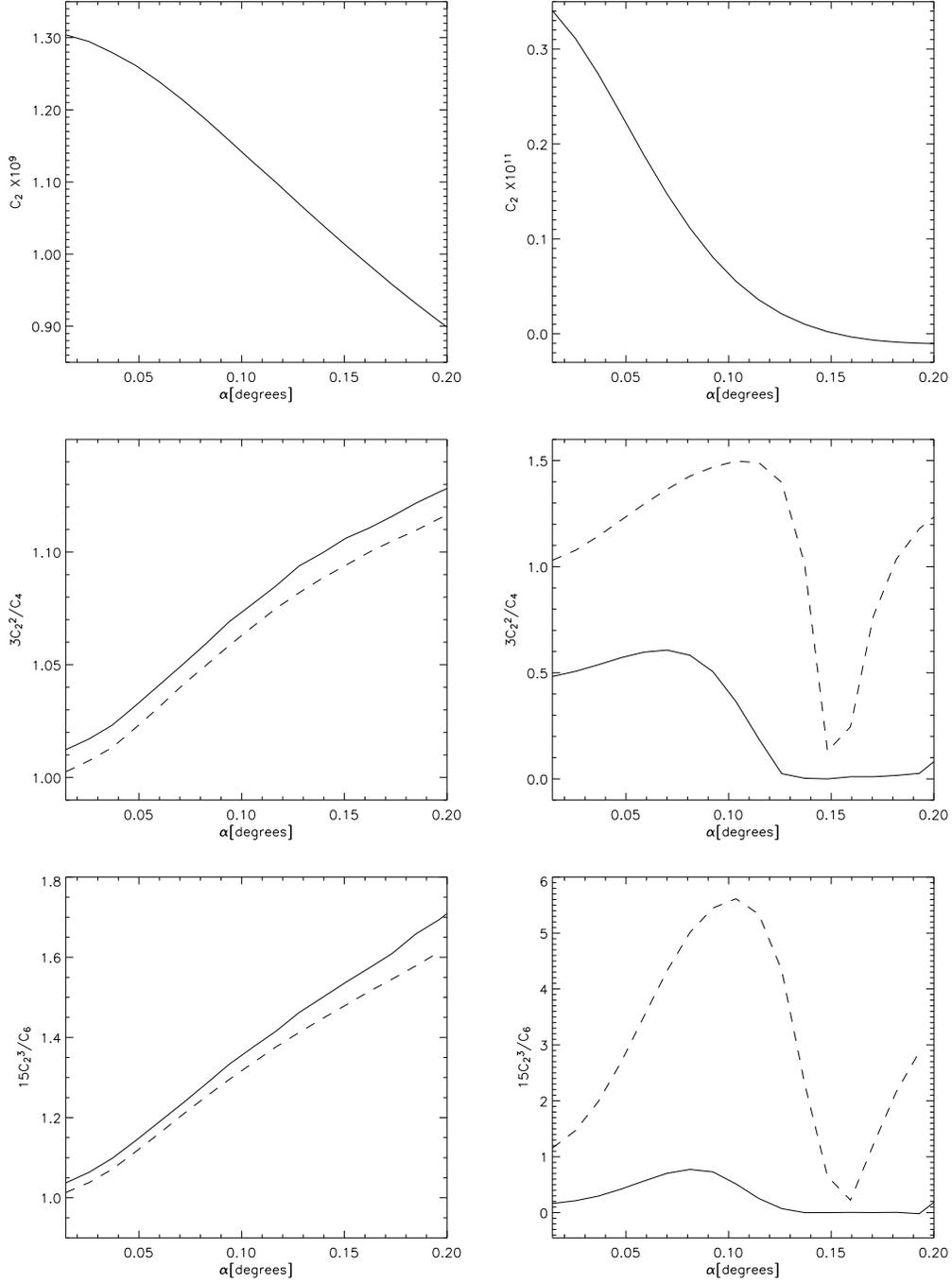}
\caption{Left (right) panels correspond to U
(D) maps.
Top, middle and bottom panels exhibit functions $C_{2}$, 
$3C^{2}_{2}(\alpha)/C_{4}(\alpha)$ and
$15C^{2}_{3}(\alpha)/C_{6}(\alpha)$
against the correlation angle $\alpha$ in 
degrees. In the solid lines, correlations 
$C_{2}$, $C_{4}$, and $C_{6}$ are directly extracted 
from the maps, whereas the ratios shown 
in the dotted lines are calculated using the 
Gaussian correlation $C_{g4}$ and $C_{g6}$ defined in
the text. The separation between continuous and dotted lines 
measure deviations from Gaussianity.
\label{fig4}}
\end{figure}

We begin with the analysis of small U and D maps of type 1. 
This study is performed with the essential aim of testing the
method designed in \S 2 to look for deviations from Gaussianity, 
this method is used in two oposite situations:
almost Gaussian U maps and fully non-Gaussian D maps.
For U maps, the question is:  can we use the method of \S 2 
to detect small deviations with respect to
Gaussianity due to both the map making method and the 
correlation analysis? The D maps have been considered 
to verify that large deviations from Gaussianity are easily detected 
and measured with the same method. For U maps, 
results are presented in the left panels of Fig.~\ref{fig4}.
Right panels show deviations from Gaussianity 
obtained from the HR-D maps. 
In this Figure, solid lines display 
correlations directly extracted from the simulated maps, 
whereas dashed lines show quantities associated to the
$C_{2}$ correlation function --extracted from the 
simulated maps-- in the case of
Gaussian statistics ($C_{g4}$ and $C_{g6}$ in Eqs. (\ref{cg4})
and (\ref{cg6})).
In the top left panel, 
the resulting $C_{2}(\alpha)$ correlation is shown 
in an appropriate $\alpha$ interval where lens
correlations are significant (see the top right panel).
The middle (bottom) panels display the ratio 
$3C^{2}_{2}(\alpha)/C_{4}(\alpha)$
($15C^{2}_{3}(\alpha)/C_{6}(\alpha)$). 
The values of these two ratios are unity for a vanishing 
correlation angle ($\alpha = 0$).
Small relative deviations 
of the dashed (Gaussian statistics) and solid (map statistics) lines 
of the left panels suggest that the U maps are rather Gaussian by
construction. A part of these deviations is associated to
the use of small maps. Using more extended maps and a 
greater total coverage,
these deviations decrease because of a smaller 
sampling variance (a better estimate of $C_{2}$, $C_{4}$, $C_{6}$, 
$C_{g4}$ and $C_{g6}$). For a large enough coverage,
the residual deviations from Gaussianity would be essentially 
due to map making (see below).
Right panels 
are the averaged correlations obtained
from the HR-D maps. These panels  
show deviations between solid and dashed lines which are much greater than those
of the corresponding left panels;
hence, D maps are not Gaussian at all.
The ratios presented in the bottom and middle panels 
are very useful to make well visible --in the figure--
the deviations from gaussianity of the U maps
(left panels). In the right panels, dashed curves present
a singular point where the $C_{2}$ correlation 
of the D maps vanishes (see the top right panel).
Very similar results are obtained from LR-D and IR-D simulations.
Once it has been verified that, even for
small type 1 maps,  the method of \S 2 works,
it can be used in other situations.

Let us now analyze L and U maps of type 2.
The differences between the correlations of the L and U maps
are hereafter called correlation increments. L maps 
(not D maps) are simplified simulations --without noise, 
contaminants and so on-- of what should be observed 
in experiments and, consequently, correlation increments
(excesses or defects with respect to the theoretical 
correlations 
in the absence of lensing) play an important role in
order to study whether lensing imprints can be observed in 
a given experiment. 
An accurate numerical calculation of the difference between 
the angular power spectra 
of L and U maps ($C_{\ell} $ increments) is not trivial. There are various 
numerical problems. In general, the 
methods designed to extract the $C_{\ell} $ coefficients from 
small maps lead to the worst results in the $\ell $ intervals where
the curvature of the angular power spectrum is great; namely,
close to its maxima and minima. The extracted $C_{\ell}$ 
coefficients deviate from their true values in these
intervals and the smaller the map the greater these deviations.
Furthermore, $C_{\ell} $ increments are differences between  
two similar angular power spectra (which are calculated with 
numerical techniques leading to small errors) and,
evidently, such a subtraction --although possible-- could
become problematic; however, 
the $C_{\ell} $ coefficients of the D maps 
(which measure correlations in maps of pure lens deformations)
are not either oscillatory quantities 
or differences between close spectra and,
consequently, they can be accurately obtained with small
maps and used to do
many theoretical and numerical studies about lensing (see below).

The increments of the $C_{\ell} $ quantities
are exhibited in Fig.~\ref{fig5}. 
As it is well known, 
the spectrum of the L maps
is very similar to that of the U maps and, consequently,
$C_{\ell} $ increments are small;
see \citet{selj96} and \citet{hu00} for previous comparisons of
L and U angular power spectra.
The dotted line of Fig.~\ref{fig5}
shows the increments given by the CMBFAST code for
the model under consideration. 
The increments of the solid line are calculated in the
following form: (i) sixteen $ 7.32^{\circ} \times 7.32^{\circ} $
U maps are simulated and, then, these maps are deformed with 
sixteen $ 7.32^{\circ} \times 7.32^{\circ} $ maps
of total lens deviations (see above) to get sixteen L maps, (i) the $C_{\ell} $
coefficients of all the U and L maps are numerically obtained;
the method used to do that was described in 
\citet{arn02} and \citet{bur03}. It is a very efficient method 
to analyze small uniformly pixelised squared maps, and (iii)
the averages of the $C_{\ell} $ quantities extracted 
from L and U maps are subtracted.
Solid and dotted lines have the same oscillatory character, but
the amplitudes of the maxima and minima are not identical,
it is mainly due to the small coverage of the analyzed maps.
In fact, it has been verified that the difference between the amplitudes
of the dotted and solid lines  
increases when our study is based on less extended maps.
Unfortunately, the use of more extended maps has a great computational 
cost and it is not necessary in this paper, where the 
comparison of the dotted and solid lines of Fig.~\ref{fig5}
strongly suggest that our numerical methods work.
Let us now estimate $C_{m}$ correlations for even $m$ values 
greater than two. It is done with the
essential aim of analyzing deviations from Gaussianity
due to lensing.

In order to estimate the $C_{m}$ correlations, the sets
of directions shown in Fig.~\ref{fig2} are placed at $N_{c}$ 
random positions 
on the map and, in this way, the average of Eq. (\ref{corr}) is performed. 
For maps with fixed size and resolution, the resulting correlations
only depend on: (1) the number $N_{c}$, and (2) the accuracy of the 
interpolation method used to get the temperature contrast
in the directions that do not point towards map nodes.
On account of these facts, U and L maps of 
$ 7.32^{\circ} \times 7.32^{\circ} $, with a resolution close to 
one arcminute, have been analyzed for different values of
$N_{c}$ and using very distinct interpolation procedures.
In this way, it has been verified that, for the two interpolations 
we have tried (bilinear and splines based on four cells), and 
for $N_{c}$ values close to  $5 \times 10^{5}$, the accuracies of
the resulting $C_{2}$, $C_{4}$ and $C_{6}$ correlations are good enough 
to allow us --using the method of \S 2-- the study and comparison 
of the deviations from 
Gaussianity in U and L maps. In order to apply that method,
fifty maps of type 2 are analyzed as follows: (a)
the averaged 
$C_{2}$, $C_{4}$ and $C_{6}$ correlations of the fifty maps 
are calculated using a certain interpolation method and 
a selected $N_{c}$ value, (b) equation (\ref{cg4}) is used to 
get $C_{g4}$ and, (c) the relative differences 
$\Delta C_{4} / C_{4}= 2(C_{4}-C_{g4})/(C_{4}+C_{g4})$ 
are calculated. This analysis is done for U and also for L maps.
We begin with the U maps.
Typical relative differences resulting from the
analysis of an unique U map are of the order of $10^{-2}$; whereas
the averaged value corresponding to a few tens of U maps is only of the 
order of $10^{-3}$. It is due to the fact that, as the number of maps 
increases, the estimation of the correlations is better, and the 
quantity $\Delta C_{4} / C_{4}$ (which vanishes in the Gaussian case)
decreases to approach its residual value due to deviations 
from Gausianity associated to map making (see above). 
By comparing the averaged values corresponding to thirty, forty and fifty
U maps, it has been verified that the $\Delta C_{4} / C_{4}$
ratio is close to $10^{-3}$ in all the cases. The 
same values of $\Delta C_{4} / C_{4}$ are obtained for the 
two interpolation methods and also for $N_{c}$ values greater
than $5 \times 10^{5}$.
The bottom panel of Fig.~\ref{fig6} describes the same study 
as in the top one, but the involved correlations are now
$C_{6}$ and $C_{g6}$. Results
are similar. 
All these results strongly suggest that  
the $\Delta C_{4} / C_{4}$ and $\Delta C_{6} / C_{6}$ averaged
values corresponding to fifty U maps 
(solid lines of the top and bottom panels of Fig.~\ref{fig6},
respectively) measure actual deviations from Gaussianity due to map making.
Fifty L maps have been obtained from the chosen U maps by using independent 
realizations of lens deviations. These maps have been analyzed in the
same way as the U maps.
It has been verified that
for thirty, forty and fifty L maps, the resulting $\Delta C_{4} / C_{4}$
and $\Delta C_{6} / C_{6}$ 
values are always around $20$ \% greater than those of the U maps. 
This percentage is
almost independent on the interpolation procedure for any  
$N_{c}$ value greater than $5 \times 10^{5}$.  
The dotted lines of the top and botom panels of Fig.~\ref{fig6} shows 
$\Delta C_{4} / C_{4}$ and $\Delta C_{6} / C_{6}$ 
relative differences obtained from 
fifty L maps. Independent lens deviations have deformed the U maps
in such a way that the $\Delta C_{4} / C_{4}$ and 
$\Delta C_{6} / C_{6}$ quantities of
the resulting L maps sistematically exceed those of the 
U maps. These quantities
measure the excess of non-Gaussianity in the L maps beyond
what is seen in the U maps.
All these results strongly suggest 
that, although deviations from Gaussianity 
produced by lensing are only $\sim 20$ \%  of those introduced in
the U maps
by the map making procedure, this $\sim 20$ \% can be pointed out 
with the method of \S 2.

Instead of numerical deviations from 
Gaussianity associated to map making, 
observation maps involve other deviations 
due to galactic contaminants, the Sunyaev-Zeldovich effect
\citep{coo01}, and so on, which are coupled to 
lens deviations from Gaussianity. Before
any eventual separation of the different 
deviations from Gaussianity 
appearing in observation maps: each contribution 
and its possible cross correlations with the remaining ones must be
characterized, including lens contribution 
we are analyzing here. 

Let us now return to the analysis of small type 1 maps with the 
essential aim of justifying that the initial 
redshift for lensing calculations ($Z_{in} = 5.2$) and the
resolution of the N-body simulations have been appropriately
chosen. In order to do that, 
we study both the contribution of the different spatial scales 
to lens deformations of the CMB and the time variation
of these contributions. 
Results are presented in Fig.~\ref{fig7}.
In order to do such a study, scales larger than 
$L_{cut} =60 \ Mpc$ (top panels), 
$L_{cut} =30 \ Mpc$ (middle panels),  and $L_{cut} = 15 \ Mpc$
(bottom panels),
have been erased in the peculiar
gravitational potential and, in each of these three cases,
the D maps and their $C_{\ell}$ quantities have been calculated 
from $z_{in} = 5.2$ to redshifts
3.9 (pointed-dashed lines), 2.6 (dashed lines), 1.3 (pointed lines)
and 0. (continuous lines).  
Left, central and right panels correspond to an 
arbitrarily chosen LR, IR and HR
n-body simulation.  
In all the panels, the angular power spectrum increases with
the redshift and it is very small at $z=3.9$; hence, for the scales 
under consideration, an initial redshift of
5.2 is a very good choice, and negligible lens deviations are 
expected at $z > 5.2$. Furthermore,
the most important part of the effect is produced between
redshifts 1.3 and 0 (compare solid and pointed lines in 
all the panels). 
In any panel (fixed cutoff and simulation),  
the maxima of the curves shift to right as the redshift 
increases. It is due to the fact that 
the angle subtended by the involved scales depends on the 
redshift. Since this angle ($\ell $) is a decreasing (increasing) 
function of the 
redshift, the maxima shift to right. 
For a given redshift, 
the curves of the middle panels are
shifted to right (left) with respect to those of the top (bottom)
panels. It is due to the fact that --as a result of the cutoff-- 
the scales of the middle panels are smaller (greater) 
than those of the top (bottom) one and, consequently, they 
subtend a smaller (greater) angle for a fixed redshift.
From the comparison of the continuous lines in the top and middle
panels it follows that spatial scales between 60 and 30 $Mpc$, whose 
effects are (are not) included in the top (middle) panel, are
responsible for the most important part of the total effect 
(solid line of the top panel).
Similarly, the comparison of the solid lines of
the middle and bottom panels indicates that the effect
produced by the scales between 30 and 15 $Mpc$ is  
greater than that produced by the scales smaller than 
15 $Mpc$ in all the cases. This last effect due to scales  
smaller than 15 $Mpc$ is very small for LR simulations (left 
bottom panel), but a comparison of the bottom panels 
--in which the n-body resolution 
increases from left to right-- among them lead to the conclusion that 
this effect of scales smaller than 15 $Mpc$ grows as the
n-body resolution increases.
It is due to the fact that the greater the n-body resolution
the greater the growing of the scales 
smaller than 15 $Mpc$.
In the case of the HR simulation of the right panels of 
Fig.~\ref{fig7}, a cutoff at 7.5 $Mpc$ has been also 
performed. Results are presented in Fig.~\ref{fig8} 
(same structure as Fig.~\ref{fig7}), 
where we see that the amplitude of the effect produced
by scales smaller than 7.5 $Mpc$ is much smaller than 
that produced by the complementary scales (up to 60 $Mpc$).
Of course, this effect should increase as resolution does,
nevertheless, for the angular scales of interest, this effect
would be negligible --for any resolution-- 
as compared with that of the remaining 
scales.
From the above considerations, it follows that the most significant 
scales are greater than $15 \ Mpc$ in all cases.
Fortunately, these scales  are rather well described by our 
LR and IR simulations and, consequently,
no important differences appear among the averaged spectra 
of our LR-D, IR-D and HR-D maps.
These spectra are presented in Fig.~\ref{fig9}, 
where we see that, for great enough $\ell $ values, 
the larger the n-body resolution the greater the 
resulting $C_{\ell}$ quantities.
These $\ell $ values correspond
to angles subtended by small enough spatial scales which 
grow more as resolution increases.
We see that LR simulations suffice to get a
very good approach to the angular power spectrum of
lens deformations.

Twenty LR simulations have been used in our study. 
Each of these 
simulations leads to eight hundred D maps which
can be analyzed to get an averaged 
angular power spectrum; in this way, 
twenty different power spectra appear whose mean 
has been represented in the solid line of Fig.~\ref{fig9}.
We are now interested in the distribution of these twenty spectra 
around the mean spectrum of Fig.~\ref{fig9}. 
How much separated are these spectra from their mean?
In order to answer this question, the mean and variance 
of the twenty $C_{\ell}$ coefficients have been 
calculated for every $\ell$ value and, afterwards,  mean, 
mean plus variance, and mean minus variance
have been used to built up solid, dotted and dashed lines
of the top panel of Fig.~\ref{fig10}, which gives a
good idea about the dispersion of the spectra associated
to each of the twenty LR simulations. 
The middle (bottom) panel of Fig.~\ref{fig10} has the same
structure as the top one, but it corresponds to
IR (HR) simulations.

Since a LR simulation suffices to get a rather good description 
of CMB lensing, we have developed a new LR simulation in a box
of $512 \ Mpc$ ($256^{3}$ cells and $256^{3}$ particles) using:
(i) a cutoff at $60 \ Mpc$, (ii) fifteen boxes to cover the 
photon trajectories from $z_{in} = 5.2$ and, (iii) a preferred direction
leading to a separation distance $L=103.8 \ Mpc$. 
Evidently, the situation is better than in previous 
simulations because of the smaller value of the number of
boxes and the larger values of $L$ and $L_{box}$.
Since the box size is twice that of previous simulations,
the new L and D maps are greater than 
those we have already analyzed, they have
$3.66^{\circ} $ degrees per edge. 
New U maps have been built up in concordance 
with the new coverage.
In the top panel of Fig.~\ref{fig11}, we show the spectra of the resulting 
D maps at
redshifts 3.9, 2.6, 1.3, and 0. (same format as in Figs.~\ref{fig7}
and ~\ref{fig8}). The spectrum is very similar to 
those obtained with previous LR simulations (compare with
the top left panel of Fig.~\ref{fig7} taking into account that
we are considering the spectra of two particular 
LR simulations). These results strongly suggest
that boxes of $256 \ Mpc$ are large enough to
study CMB lensing.

Results obtained from Eq. (\ref{devi}),
in position space, are now compared with those obtained,
in momentum space, from
Eqs. (\ref{funda1}) and (\ref{funda2})
(plus statistics based on $128^{3} $ observers,
see \S 2). In order to do this comparison,
LR simulations as 
those used along the paper (excepting previous
paragraph) are considered. The cutoff is
performed at $60 \ Mpc$.
The main difference with respect to previous cases
is the size and resolution of the simulated D maps.
We have not been able to work with $128 \times 128$
directions --in momentum space-- as a result of the 
high computational cost of the method based
on Eqs. (\ref{funda1}) and (\ref{funda2}), which works 
with $128^{3}$ observers at the same time.
By this reason and with the essential aim of 
comparing results in position and momentum 
spaces, regular maps covered
by $16 \times 16$ directions with a resolution of 
$5^{\prime} $ and an edge of $1.25^{\circ} $
have been simulated. Size and
resolution are smaller than in previous cases and,
consequently, only 
some $C_{\ell} $ quantities 
can be extracted from the new D maps with admissible
accuracy. In any case, the D maps can be simulated 
using the two methods and identical U maps;
afterwards, the $C_{\ell} $ coefficients 
can be extracted using the same codes and, finally, 
results can be compared.
The solid line of the bottom panel of Fig.~\ref{fig11}
shows the spectrum obtained, in position 
space, by averaging the spectra of twenty D maps
with the new size and resolution. 
Dashed and dotted lines exhibit the
$C_{\ell}$ quantities found, in momentum space, using only
one LR simulation in each case, but considering 
$128^{3}$ observers in the 
statistical analysis (see \citet{cer04} and \S 2). 
Figure~\ref{fig11} shows that both methods lead to 
similar curves.
If we average over twenty simulations in momentum space,
the final spectrum is close to that 
displayed in the solid line of Fig.~\ref{fig11}; namely,
both methods lead to similar final spectra separately.

\section{Discussion and conclusions}

A method for ray-tracing through n-body PM simulations 
\citep{cer04} is used to 
simulate lensed maps of the CMB. Our attention is focused on the study of
deviations from Gaussianity in these maps. Some correlation functions 
as $C_{4} $, $C_{6} $,
$C_{g4} $ and $C_{g6} $ are estimated and compared --for the first 
time-- to measure deviations from Gaussianity. Considerable emphasis 
is put on the comparison of the lens deformations 
produced by different small spatial scales, and also in the 
study of the evolution of these deformations.

For a given map, the relative difference between
the correlation functions $C_{4}$ and $C_{g4}$,
and also between $C_{6}$ and $C_{g6}$ measure deviations from Gaussianity.
These relative differences --$\Delta C_{4}/C_{4}$ and $\Delta C_{6}/C_{6}$--
have been calculated for U and 
L maps. According to our expectations, the functions 
$\Delta C_{4}/C_{4}$ and $\Delta C_{6}/C_{6}$
of the U maps take on small values pointing out 
small deviations from Gaussianity
due to construction (see \S 3).
This success in the analysis of U maps strongly suggest that the 
same method
could be applicable to the analysis of other deviations from Gaussianity
(lensing, Sunyaev-Zelovich and so on). 
In the case of L maps, quantities $\Delta C_{4}/C_{4}$ 
takes on values which are 
greater than the values of the U maps. The same occurs for
$\Delta C_{6}/C_{6}$.
The differences 
between the $\Delta C_{4}/C_{4}$  
($\Delta C_{6}/C_{6}$)  quantities of the L and U maps 
are about the twenty per cent of the
$\Delta C_{4}/C_{4}$  
($\Delta C_{6}/C_{6}$) values corresponding to the U maps.
The same values of these differences are obtained 
for different interpolation procedures and whatever 
the value of the parameter $N_{c} > 5 \times 10^{5}$ may be; hence,
these differences measure actual deviations from Gaussianity 
produced by lensing (see discussion in \S 3).

\begin{figure}
\epsscale{0.4}
\plotone{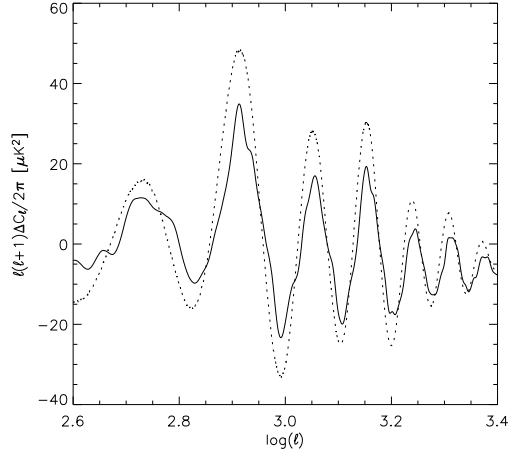}
\caption{Difference $\Delta C_{\ell}$ in $\mu K^{2}$, 
multiplied by the factor $\ell (\ell +1)/2\pi$,
{\it v.s.} $\log \ell$.
Difference is calculated between the $C_{\ell} $
coefficients of L and U maps. Dotted line has been 
obtained with CMBFAST and solid line 
from the numerical analysis of sixteen
$7.32^{\circ} \times 7.32^{\circ}$ L and U maps (see the text).
\label{fig5}}
\end{figure}

\begin{figure}
\epsscale{0.4}
\plotone{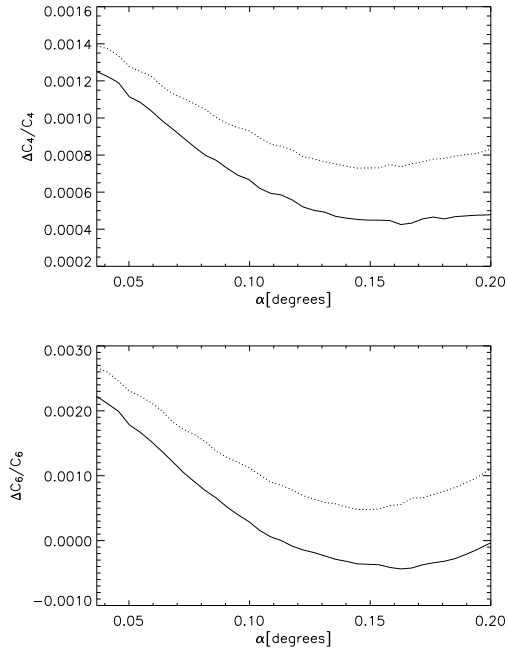}
\caption{Top panel shows relative differences $\Delta C_{4}/C_{4}$ 
between the $C_{4}$ correlations extracted from the U maps and 
the $C_{g4}$ Gaussian correlations (see text), {\it v.s.} the 
correlation angle in degrees.  
Bottom panel has the same structure as the top one, but it 
exhibits relative differences $\Delta C_{6}/C_{6}$.
\label{fig6}}
\end{figure}

\begin{figure}
\epsscale{0.8}
\plotone{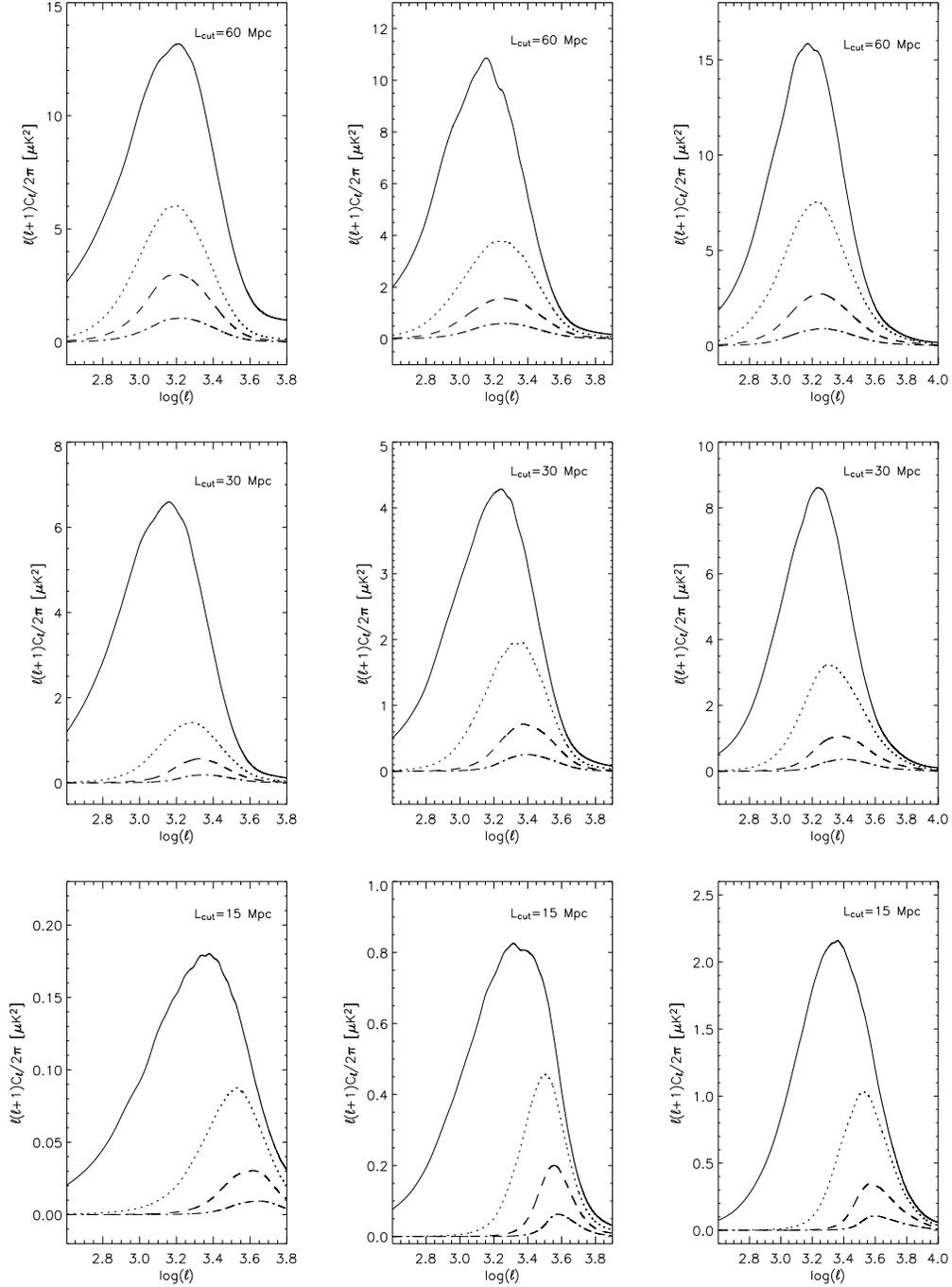}
\caption{Quantity $\ell (\ell+1) C_{\ell }/ 2 \pi$ in $\mu K^{2}$
{\it v.s.} $\log \ell$. Left, central and right panels involve
LR, IR and HR simulations. Top, middle and bottom panels 
correspond to $L_{cut}$ values of $60$, $30$, and $15 \ Mpc$. 
Solid, pointed, dashed, and pointed-dashed lines
display the $C_{\ell}$ coefficients of the D maps produced 
by lensing between
$z_{in}=5.2$ and the redshifts $3.9$, $2.6$, $1.3$, and $0.$,
respectively.  
\label{fig7}}
\end{figure}

\begin{figure}
\epsscale{0.4}
\plotone{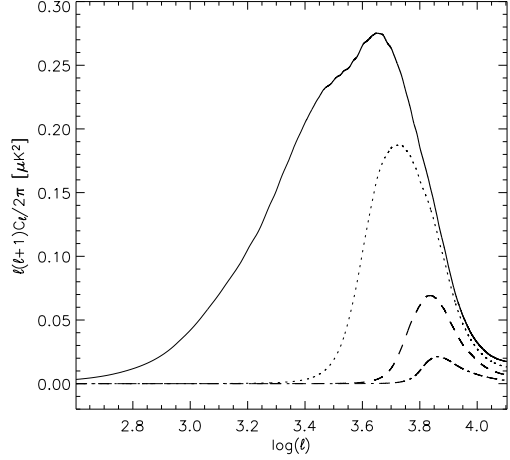}
\caption{Same as in Fig.~\ref{fig7} for the HR simulation 
of that Figure (right panels), 
with a cutoff at $L_{cut}=7.5 \ Mpc $ 
\label{fig8}}
\end{figure}

\begin{figure}
\epsscale{0.4}
\plotone{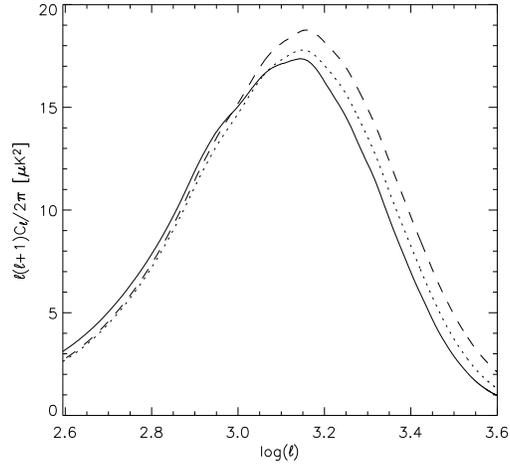}
\caption{Quantity $\ell (\ell+1) C_{\ell }/ 2 \pi$ in $\mu K^{2}$               
{\it v.s.} $\log \ell$. The $C_{\ell }$ coefficients involved 
in the solid, dotted and dashed lines correspond to 
the averaged power spectra of the sets of maps LR-D, IR-D and HR-D
defined in the text
\label{fig9}}
\end{figure}

\begin{figure}
\epsscale{0.4}
\plotone{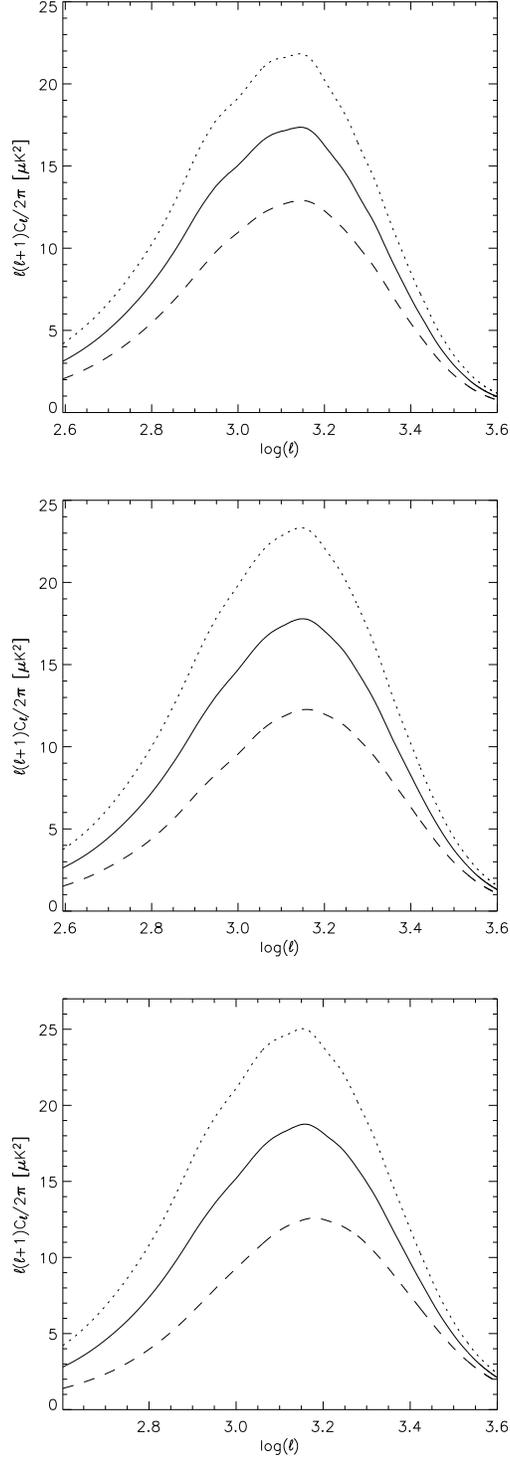}
\caption{Quantity $\ell (\ell+1) C_{\ell }/ 2 \pi$ in $\mu K^{2}$               
{\it v.s.} $\log \ell$. Top, middle and bottom panels show results
from LR, IR and HR simulations. Solid lines are the averaged 
spectra displayed in Fig.~\ref{fig9}, whereas the dotted and dashed 
lines show $1 \sigma $ curves in the spectra distribution
(see the text for more details).
\label{fig10}}
\end{figure}

\begin{figure}
\epsscale{0.4}
\plotone{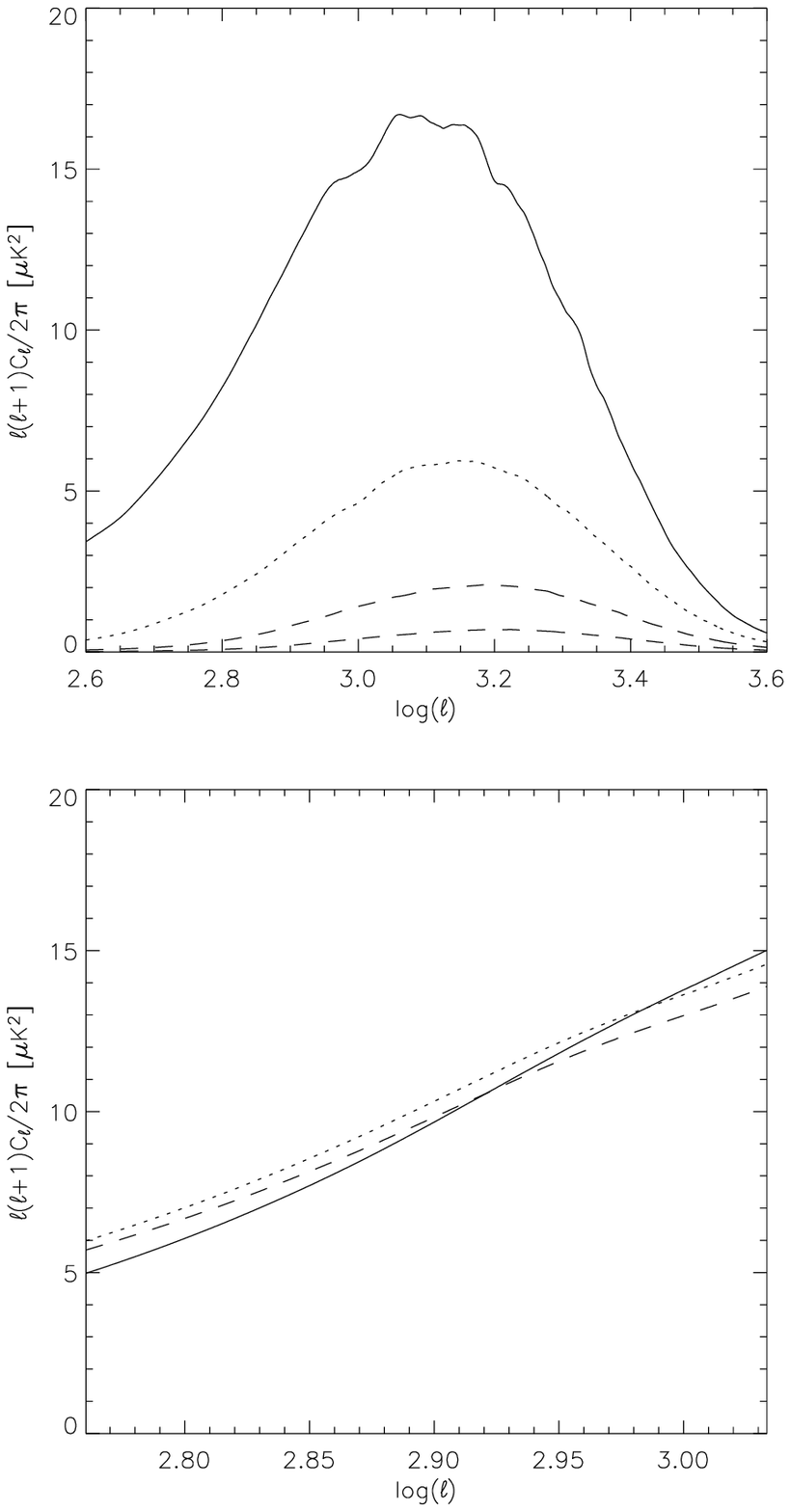}
\caption{Top panel has the same structure as panels of Fig~\ref{fig7}.
It corresponds to a LR simulaton in a big box of $512 \ Mpc $,
with a cutoff at $L_{cut} = 60 \ Mpc$. Bottom panel
represents the same function as in the top one, but the $C_{\ell }$
coefficients have been obtained from D maps with a size of $1.25^{\circ}$
and a resolution of $5^{\prime}$. 
The $C_{\ell }$ quantities of the solid line
correspond to a map obtained from Eq. (\ref{devi}) in position space,
whereas in the dotted and dashed lines, 
we present the $C_{\ell }$ coefficients of a map, which has been 
found from Eqs. (\ref{funda1}) and (\ref{funda2})
in momentum space (see the text for more explanations). 
\label{fig11}}
\end{figure}

L and U maps lead to very similar
$C_{\ell}$ quantities. Differences
between these quantities are 
calculated with CMBFAST and also from the analysis
of our simulated maps. These differences 
are displayed in Fig. (\ref{fig5}). Both methods lead 
to comparable results. This fact strongly support  
the ray-tracing procedure, the numerical simulations, and the
correlation analysis used in this paper.

Box sizes of $\sim 256 \ Mpc$ seem to be appropriate for ray-tracing
through n-body simulations. Much greater (smaller) sizes lead to problems 
with resolution (L values, number of boxes and so on). In these boxes,
scales between $15$ and $60 \ Mpc$ --small enough as compared with 
chosen box size--
produce the largest part of 
the lens effect on the CMB, whereas scales
smaller than $15 \ Mpc$ only produces a small part.
By this reason, low resolution n-body simulations
suffice to get a good approximation to the angular correlations 
produced by lensing. For scales smaller than $60 \ Mpc$,
the most important
part of the lens effect
is produced between redshifts 2.6 and 0., whereas only a small part of
this effect is produced at $z > 3.9$.

Two independent codes based on distinct 
approaches have been used to deal with the same problem:
the estimate of lens correlations produced by scales smaller 
than $L_{cut}$.
One of these codes works in momentum space and involves
many observers \citep{cer04} for statistical analysis. 
In the second code, calculations
are performed in position space 
--method applied along this paper-- and the observer is unique.
Both codes lead to very similar angular power spectra
after averages.

Let us now list some prospects. It would be worthwhile: 
(1) the use of other n-body codes 
(allowing more resolution) with the essential aim of getting more accurate
correlations at small angular scales, (2)
the application of appropriate modern methods (wavelets, Minkowski 
functionals and so on) to study deviations from Gausianity in the L and U maps
we have simulated,
(3) the characterization of all the deviations from
Gaussianity which appear coupled in observation maps, e.g.
deviations due to the Sunyaev-Zel'dovich effect, to
the radiation from the galaxy in the CMB frequencies and so on.
The necessary statistical analysis could be developed by
using either the method of this paper or other methods (see above),
and (5) the study of the possible detection of deviations from 
Gaussianity in maps of the PLANCK mission.


\acknowledgments

This work has been                        
supported by the Spanish MCyT
(project AYA2003-08739-C02-02 partially funded with FEDER) and
also by the Generalitat Valenciana (grupos03/170).  
PC-D thanks the Ministry of Science and Technology for a fellowship.
Calculations were carried out at the 
Centro de Inform\'atica de la Universidad de Valencia 
(CERCA and CESAR).

\end{document}